\documentclass[11pt,twoside]{article}
\usepackage{asp2010}

\resetcounters

\begin{document}

\title{Prototype Implementation of Web and Desktop Applications for ALMA Science Verification Data and the Lessons Learned}
\author{Satoshi~Eguchi$^1$, Wataru~Kawasaki$^1$, Yuji~Shirasaki$^1$, Yutaka~Komiya$^1$, George~Kosugi$^1$, Masatoshi~Ohishi$^1$, and Yoshihiko~Mizumoto$^1$
\affil{$^1$National Astronomical Observatory of Japan, 2-21-1, Osawa, Mitaka, Tokyo, 181-8588 Japan}}

\begin{abstract}
ALMA is estimated to generate TB scale data during only one observation; astronomers manage to identify which part of the data they are really interested in.
Now we have been developing new GUI software for this purpose utilizing the VO interface: ALMA Web Quick Look System (ALMAWebQL) and ALMA Desktop Application (Vissage).
The former is written in JavaScript and HTML5 generated from Java codes by Google Web Toolkit, and the latter is in pure Java.
An essential point of our approach is how to reduce network traffic: we prepare, in advance, "compressed" FITS files of 2x2x1 (horizontal, vertical, and spectral directions, respectively) binning, 2x2x2 binning, 4x4x2 binning data, and so on.
These files are hidden from users, and Web QL automatically choose proper one by each user operation.
Through this work, we find that network traffic in our system is still a bottleneck towards TB scale data distribution.
Hence we have to develop alternative data containers for much faster data processing.
In this paper, I introduce our data analysis systems, and describe what we learned through the development.
\end{abstract}

\section{Introduction}

Atacama Large Millimeter/submillimeter Array (ALMA) is the largest radio telescope
built on the Chajnantor plateau in northern Chile, and it is expected to provide us
much useful information for deep understanding of the universe due to its unprecedented
high resolutional data in space and spectrum.

ALMA is estimated to generate $\sim$200\ TB observational raw data every year,
and the volume of a processed data cube for one target may exceed $\gtrsim$2\ TB.
One has to consider 1) how to bring such a big data cube into his/her computer,
and 2) how to extract fruitful information from the data cube on the computer.

\section{Method}

To solve these problems, we apply ``binning'' and ``cut-out'' to a data cube.
An ALMA data cube has 4 dimensions: right ascension (R.A.), declination (Dec.), channel
(frequency), and polarization.
Current Science Verification data do not have polarization, and they are
identical to 3-dimensional data cubes.
In the binning stage, we make every $q \ (q = 1, 2, 2^{2}, 2^{3}, \cdots)$ pixels into one pixel
in the R.A. and Dec. directions, and every $p \ (p = 1, 2, 2^{2}, 2^{3}, \cdots)$ channels into
one channel.
By the binnings, the data cube size becomes $p q^{2}$ times smaller than original one.
We prepare all binning data cube in advance.
In the cut-out stage, one should download a subset of the data cube where he/she is
interested in.
By a combination of binning and cut-out, we can reduce network traffic drastically.

\section{System Configuration and Use Case}

\begin{figure}
	\epsscale{0.85}
	\plotone{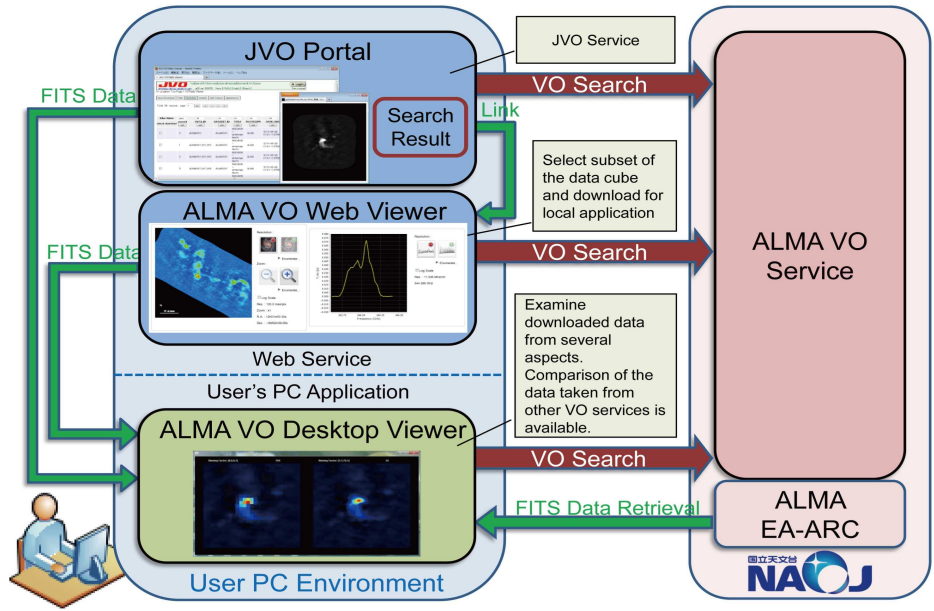}
	\caption{A perspective of the system. There are two web and one desktop applications.
	\label{O10_f1}}
\end{figure}

We developed a new on-line system \citep{D5_adassxxii} consists of two web applications
(JVO portal and web viewer) and one desktop application
(desktop viewer; \citealt{P047_adassxxii}) shown in Figure~\ref{O10_f1}.

The use case is below:
\begin{enumerate}
	\item A user visits Japanese Virtual Observatory (JVO) portal site:
	\url{http://jvo.nao.ac.jp/portal/}.
	In this site, the user searches and selects data cubes which
	contains his/her interested objects.
	However, he/she cannot figure out where his/her interested region
	in each data cube is, and which resolutions are adequate for his/her
	research.
	
	\item The user launches the web viewer from the result pages.
	The web viewer enables the user to change resolutions, viewing position, and
	zoom graphically, hence he/she confirms the desired cut-out region and binning
	factors exactly.
	
	\item The user download the data cubes to his/her local hard disk via the
	web viewer, and launches the desktop viewer.
	
	\item The desktop viewer has much functionality to look at each data cube in detail,
	and the user can obtain enough information about the objects by this viewer.
\end{enumerate}
The user can iterate this process if he/she wants.

\begin{figure}
	\epsscale{0.85}
	\plotone{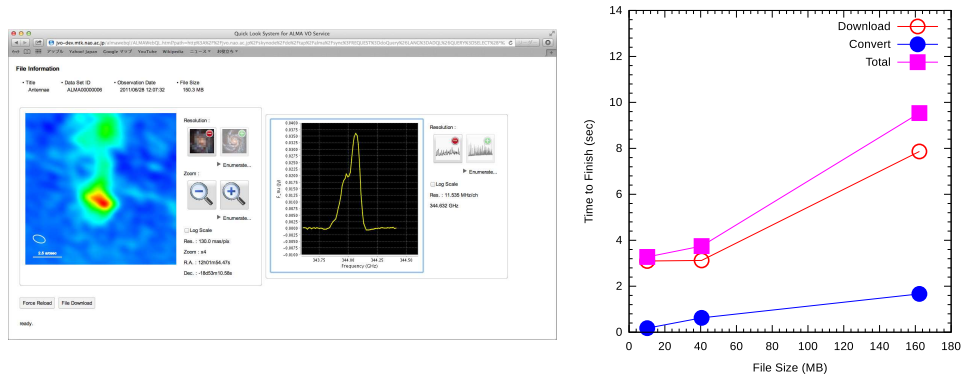}
	\caption{{\bf Left:} a screen shot of ALMAWebQL. With the web application, one can change resolutions,
	viewing position, and zoom on a web browser.
	{\bf Right:} the relation between fits file size and processing time. Blue filled circles, red
	open circles, and magenta filled rectangles represent file conversion time, file downloading time,
	total processing time, respectively.\label{O10_f2}}
\end{figure}

\section{ALMAWebQL}

\subsection{Implementation}

ALMAWebQL is a web viewer which enables a user to find his/her sufficient binning
factors and a cut-out region graphically (Figure~\ref{O10_f2}, left)\footnote{
You can watch the screen video on \url{http://youtu.be/48IJ5G-5cG4}.}.
This is an Ajax application built on Google Web Toolkit, utilizing the HTML5 Canvas component.

A web browser provides very limited programming interfaces, and it is difficult to access to
the local file system inside the browser.
However, one can perform simple operations without any servers by utilizing  JavaScript.
I let a web browser do as many, {\bf but simple} operations, as possible to reduce network
traffic.
In case of ALMAWebQL, I implemented coordinate transforms and zoom functionality by
JavaScript.

\begin{figure}[tbh]
	\begin{center}
		\includegraphics[keepaspectratio,width=0.54\hsize,clip]{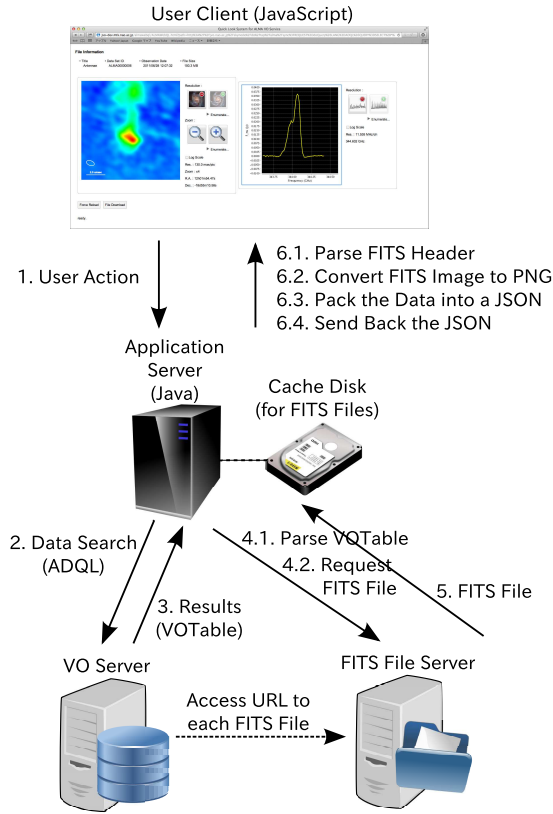}
	\end{center}
	\caption{A schematic block diagram of ALMAWebQL. The application consists of
	a user client written in JavaScript and an application server written in Java. Any user actions
	are translated into ADQL and the results are translated into a JavaScript object on the application
	server.\label{O10_f3}}
\end{figure}

Figure~\ref{O10_f3} is a schematic block diagram of ALMAWebQL, which consists of a web
client and an application server.
The application server contacts the VO server in ADQL, and retrieves a proper FITS file
from the file server when a user does some actions on the web client.
Then the application server sends back the FITS header information and image to the
web client as a JavaScript object.

\subsection{Problems towards a Production Release}

I evaluate the system performance and find some problems towards a production release.
Figure~\ref{O10_f2} (right) shows the relation between FITS file size and file processing time
of the application server.
It is obvious that file retrieving time is dominating.
To solve the problem, the file server should be more intelligent to reduce network
traffic.

Currently, this is only on an idea stage, but a new database savvy hierarchical data
cube container might be a solution.

\section{Summary}

ALMA generates a big data cube for one object, and it may exceed 2\ TB.
On the other hand, both network bandwidth and computer resource are limited.
To solve these problems, we developed a VO system consists of two web applications
and one desktop application.
The aim of this system is to reduce network traffic by binning and cut-out.

We provide a web viewer to let a user select his/her adequate binning factors
and cut out his/her interested region visually, and also provide a desktop viewer
to perform detailed quick look.
However, network traffic is still a bottleneck towards a production release,
and a more intelligent file server is required.


\bibliography{O10}

\end{document}